\def\kms{\ifmmode{\,\hbox{km}\,s^{-1}}\else {\rm\,km\,s$^{-1}$}\fi}

\def\msun{{\rm\,M_\odot}}

\def\kmsm{{\rm\,km\,s^{-1}\,Mpc^{-1}}}

\def\hmpc{\ifmmode{h^{-1}\,\hbox{Mpc}}\else{$h^{-1}$\thinspace Mpc}\fi}

\def\eg{{\it e.g.}~}

\def\et{{\it et~al.}~}

\def\ie{{\it i.e.}~}
\def\sigp{\ifmmode{\sigma_p}\else {$\sigma_p$}\fi}
\def\sig1{\ifmmode{\sigma_1}\else {$\sigma_1$}\fi}
\def\r200{\ifmmode{r_{200}}\else {$r_{200}$}\fi}
\def\m200{\ifmmode{M_{200}}\else {$M_{200}$}\fi}

\documentstyle[11pt,aaspp4]{article}
\tighten

\begin{document}

\title
{The Average Mass Profile of Galaxy Clusters}

\author
{
R.~G.~Carlberg\altaffilmark{1,2},
H.~K.~C.~Yee\altaffilmark{1,2},
E.~Ellingson\altaffilmark{1,3},
S.~L.~Morris\altaffilmark{1,4},
R.~Abraham\altaffilmark{1,4,5},
P.~Gravel\altaffilmark{1,2},
C.~J.~Pritchet\altaffilmark{1,6},
T.~Smecker-Hane\altaffilmark{1,4,7}
F.~D.~A.~Hartwick\altaffilmark{6}
J.~E.~Hesser\altaffilmark{4},
J.~B.~Hutchings\altaffilmark{4},
\& J.~B.~Oke\altaffilmark{4}
}

\altaffiltext{1}{Visiting Astronomer, Canada--France--Hawaii Telescope, 
        which is operated by the National Research Council of Canada,
        le Centre National de Recherche Scientifique, and the University of
        Hawaii.}
\altaffiltext{2}{Department of Astronomy, University of Toronto, 
        Toronto ON, M5S~3H8 Canada}
\altaffiltext{3}{Center for Astrophysics \& Space Astronomy,
        University of Colorado, CO 80309, USA}
\altaffiltext{4}{
        National Research Council of Canada,
        Herzberg Institute of Astrophysics,     
	Dominion Astrophysical Observatory, 
        5071 West Saanich Road,
        Victoria, BC, V8X~4M6, Canada}
\altaffiltext{5}{Institute of Astronomy, 
        Madingley Road, Cambridge CB3~0HA, UK}
\altaffiltext{6}{Department of Physics \& Astronomy,
        University of Victoria,
        Victoria, BC, V8W~3P6, Canada}
\altaffiltext{7}{Department of Physics \& Astronomy,
        University of California, Irvine,
        CA 92717, USA}

%\clearpage

\begin{abstract}
The average mass density profile measured in the CNOC cluster survey
is well described with the analytic form
$\rho(r)=Ar^{-1}(r+a_\rho)^{-2}$, as advocated on the basis on n-body
simulations by Navarro, Frenk \& White. The predicted core radii are
$a_\rho=0.20$ (in units of the radius where the mean interior density
is 200 times the critical density) for an $\Omega=0.2$ open CDM model,
or $a_\rho=0.26$ for a flat $\Omega=0.2$ model, with little dependence
on other cosmological parameters for simulations normalized to the
observed cluster abundance.  The dynamically derived local
mass-to-light ratio, which has little radial variation, converts the
observed light profile to a mass profile. We find that the scale
radius of the mass distribution, $0.20\le a_\rho\le 0.30$ (depending
on modeling details, with a 95\% confidence range of $0.12-0.50$), is
completely consistent with the predicted values.  Moreover, the
profiles and total masses of the clusters as individuals can be
acceptably predicted from the cluster RMS line-of-sight velocity
dispersion alone.  This is strong support of the hierarchical
clustering theory for the formation of galaxy clusters in a cool,
collisionless, dark matter dominated universe.
\end{abstract}

\keywords{galaxies: clusters, cosmology: large-scale structure of universe}

\clearpage
\section{Introduction}

A fundamental prediction of the hierarchical clustering paradigm in a
collisionless dark matter dominated universe is that the average mass
profile of dark halos must primarily depend on mass, if the density
perturbation spectrum is nearly a power law.  The single parameter
surface brightness profile, $\Sigma\propto \exp{(-r^{1/4})}$, was
proposed by deVaucouleurs (1948) to describe elliptical galaxies, but
it was later found also to be quite an accurate description of
isolated n-body collapse simulations, both with and without
dissipation (\cite{vanA,cln}). This $r^{1/4}$ ``law'' has no simple
analytic form to describe the volume distribution (\cite{young}),
motivating Hernquist (1990) to propose the function
$\rho_H(r)=Ar^{-1}(r+a)^{-3}$ as a more tractable alternative, which in
projection is usefully close to the $r^{1/4}$ function.

In cosmological halo formation simulations, where external tides and
continuing infall are included, the central density profile is found
to remain approximately $r^{-1}$ (\cite{dc,vbias,sw}), although there
is growing evidence that the central slope is somewhat steeper
(\cite{makino,ec}).  However, beginning near the virialization radius
the density profile is found to fall somewhat less steeply than in
isolated collapses, leading Navarro, Frenk \& White (1996 and 1997,
hereafter NFW) to advocate $\rho(r)=Ar^{-1}(r+a)^{-2}$ as an empirical
description of the halo density profiles.  In a series of exhaustive
n-body simulations the relation between the scale radius, $a$, of the
halo and its mass has been studied to yield an empirical understanding
of its value (\cite{cl,nfw2}).  The principal result of interest here
is that the scale radius, $a$, is measured in the simulations to be
about 20\% of the ``virial radius'' for the $10^{15}\msun$ dark halos
appropriate to rich clusters.  This result has only a weak dependence
on either the power spectrum, or the world model parameters $\Omega$
and $\Lambda$, for simulations normalized to the observed cluster
abundance.  Halos of much lower mass, say the $10^{12}\msun$
characteristic of individual galaxies, are generally much more
centrally concentrated than clusters, with quite a strong dependence
on the cosmological details.

In the next section we fit the galaxy number density profile,
$\nu(r)$, to the NFW function, then present evidence that this profile
(specifically for galaxies selected in Gunn $r$, which as a red pass
band is relatively insensitive to the current star formation rate)
accurately traces the average cluster mass profile, $\rho(r)$. In
particular we compare the derived surface mass profile with the
surface galaxy number profile.  Our analysis is independent of the
light-traces-mass assumption and allows for a range of velocity
dispersion profiles. We have previously shown that this type of
analysis works to recover the same mass profile from two vastly
different subsamples of the data (\cite{br}).  The fitted scale radius
of the NFW model is compared to the predicted scale radius. In Section
3, we use the scaling from velocity dispersion to radius to predict
the form of the profile for each cluster in turn and compare it to the
available data. The results are briefly discussed in the final
section.  We use $H_0=100\kmsm$ and $\Omega_0=2q_0=0.2$ for all our
calculations, although the results are not very sensitive to these
choices.

\section{The Average Mass Profile of a Galaxy Cluster}

The Canadian Network for Observational Cosmology (CNOC) Cluster Survey
(\cite{cnoc1,yec}) obtained $\sim2600$ redshifts of Gunn $r$ selected
galaxies in the fields of 16 high luminosity X-ray clusters at
$z\sim\onethird$. This is a relatively homogeneous sample of clusters
that are guaranteed to be at least partially virialized on the basis
of their X-ray emission.  The virial mass-to-light ratios of these
clusters were found to be identical, within their measurement errors
of $\sim25\%$ (\cite{global}). The 14 clusters that are ``non-binary''
are combined in normalized co-ordinates to make an ensemble cluster.
This diminishes substructure and asphericity to a level where the
galaxies can be treated as if in a spherical distribution which is
consistent with dynamical equilibrium (\cite{br}).  Galaxies of all
colors above a k-corrected Gunn $r$ absolute magnitude of $-18.5$ are
included in the average.

\subsection{The Average Cluster Number Density Profile}

To combine the clusters the brightest cluster galaxy (BCG) is used as
the nominal center of each cluster on the sky following our earlier
procedures (Carlberg, Yee \& Ellingson 1997, hereafter
\cite{profile}).  The galaxy velocities are normalized to
$\sigma_1$, the RMS velocity dispersion of the clusters about the
cluster mean.  The projected radii are normalized to an empirically
determined \r200, the radius where the mean interior overdensity is
$200\rho_c$.  To derive \r200\ from the observational virial radius,
$r_v$ (which is largely fixed by the outer boundary of the sample), we
assume that $M(r)\propto r$. This gives
\begin{equation}
r_{200} = {\sqrt{3}\over 10} {\sigma_1\over{H(z)}},
\label{eq:r200}
\end{equation}
which is completely independent of the observational virial radius.
Usually the extrapolation in radius is a modest 25\%, so the precise
mass profile assumed makes little difference to the result
(\cite{profile}).

The average projected galaxy number density profile, $\Sigma(R)$, is
fit with the projection of the volume density function,
\begin{equation}
\nu(r) = {A \over{r(r+a_\nu)^p}},
\label{eq:nu}
\end{equation}
where $p$ is fixed at either $p=2$ (NFW) or $p=3$ (Hernquist).  The
results for $p=2$ are shown in Figure~\ref{fig:surf} as the solid
line. The dashed lines are described below.  Both the $p=2$ and $p=3$
forms are statistically acceptable fits by the $\chi^2$ test.  We will
only consider the $p=2$ form for the rest of this paper. The fitted
scale radius is $a_\nu=0.27$ with a 95\% confidence range of
$0.13-0.43$ from the $\chi^2$ distribution.

\subsection{The Relationship between the Mass and Number Density Profiles}

There is no dynamical necessity for $\nu(r)$ to be directly
proportional to $\rho(r)$. The relationship between the two is derived
from the projected velocity dispersion, $\sigma_p(R)$. The dynamical
mass profile, $M_D(r)$, which is the volume integral of $\rho(r)$, is
inferred from the Jeans Equation (\eg, \cite{bt,profile}),
\begin{equation}
M_D(r) = -{\sigma_r^2r\over G}
        \left[{{d \ln{\sigma_r^2}}\over{d\ln{r}}} +
        {{d\ln{\nu}}\over{{d\ln{r}}}} +2\beta\right],
\label{eq:jeans}
\end{equation}
where the velocity anisotropy parameter, $\beta = 1 -
\sigma_\theta^2/\sigma_r^2$. 
N-body simulations for a variety of cosmologies show that the 
dependence of $\beta$ has a nearly universal radial variation
(\cite{cl}) which we somewhat arbitrarily model as,
\begin{equation}
\beta(r) = \beta_m {{4r}\over{r^2+4}},
\label{eq:betar}
\end{equation}
which takes on a maximum value of $\beta_m$ at $r=2$, in \r200\
units. We demonstrate below that the volume {\em mass} density is not
very sensitive to the details of the velocity modeling, provided the
model is an acceptable statistical description of the data.  We use
values of 0.3 and 0.5 for the parameter $\beta_m$ which roughly
bracket the range of $\beta$ seen in the simulations.

We model the radial velocity dispersion as,
\begin{equation}
\sigma_r^2 = B {{c_1r/(1+c_1r) + c_2}\over { 1 + r/b}},
\label{eq:sig}
\end{equation}
where $B$ and $b$ are the two parameters adjusted to fit the observed
$\sigp(R)$. We include here a new central data point at 0.05\r200.
The rest of the data are as in
\cite{profile}.  The $[c_1,c_2]$ parameters are
externally fixed to allow us to vary the shape of the curve. In
Figure~\ref{fig:sig} we display results where $[c_1,c_2]$ are $[0,1]$
and $[1,0]$ to give the two extremes of the functional form. The
$[0,1]$ form was used in
\cite{profile} and $[1,0]$ gives a dispersion which
drops to zero at the center. The intermediate values $[8,1/2]$ are
adopted for reasons made clear below. Both the $[1,0]$ and $[8,1/2]$
forms have reduced $\chi^2$ near unity. The $[0,1]$ form is about 4\%
probable, although this is critically dependent on the $0.05\r200$
point, the concern being that the old central galaxies may be a
distinct, low velocity dispersion, cluster population.

We show the results of the mass analysis in Figure~\ref{fig:dol}. The
quantity plotted is $\rho(r)/\nu(r)$ normalized to $M_v/L \times
L(r)$, computed from the same data.  The ratio $\rho(r)/\nu(r)$ is
only weakly constrained inside $0.1\r200$ and the assumption of
virialization likely fails outside $1.5\r200$. Within this radial
range $\rho(r)/\nu(r)$ varies less than about 30\% of its mean value.
It should be noted that for most purposes the more relevant quantity
is $M_D(r)$, the volume integral of $\rho(r)$, which is 
substantially more slowly varying. The offset of $\rho(r)/\nu(r)$
below unity indicates that mass and light are similarly distributed
but that the virial mass has a scale error.  A wide range of positive
intermediate values of $c_1$ and $c_2$ gives nearly constant $b_{Mv}$
everywhere. We display the $c_1=8$, $c_2=1/2$ case, which is one of
many intermediate sets of parameters which leads to substantial
cancellation of the opposite behaviour of the two extreme
$\sigma_r(r)$ models in the derived $M_D(r)$.  

With the dynamically derived $\rho(r)/\nu(r)$ in hand we can use it to
model the galaxy number profile, $\nu(r)$, assuming the {\em mass}
profile is described by the NFW profile. This procedure measures
$a_\rho$, the scale radius of the {\em mass} distribution.  We take
$\rho(r)/\nu(r)$ to be equal to the endpoint of the range for
$r<0.1\r200$ and $r>1.5\r200$.  The results for our three velocity
dispersion functions and the two values of $\beta_m$ are shown in
Figure~\ref{fig:surf} as the dashed lines.  The resulting $a_\rho$
values range from $0.20$ ($[c_1,c_2]=[0,1]$) to $0.30$ ($[1,0]$). This expands
the 95\% confidence range of $a_\rho$ over $a_\nu$ slightly, to
$0.12\le a_\rho \le 0.50$. We conclude that galaxies selected from
their red light brightness,
\ie, a passband well to the red of the 4000\AA\ break to minimize
sensitivity to star formation (and not necessarily galaxies red in
color,
\cite{br}), accurately trace the mass profile of the cluster. That
is, $\nu(r)$ is directly proportional to $\rho(r)$, within the
statistical limits of our measurements.

\section{Individual Cluster Profiles}

The NFW profile can be tested for its application to the clusters as
individuals. The
\r200\ for each cluster is derived from the observed velocity dispersion,
Equation~\ref{eq:r200}. The NFW scale radius is set at our best fit
value, $a_\nu=0.27$. The fits are restricted to the region inside
$1.5\r200$, which is expected to be virialized for $\Omega\simeq0.2$.
The resulting $\chi^2$ per degree of freedom (there being 10-16
degrees of freedom) are given in Table~\ref{tab:chi}. The cluster
MS0906+11, which we observed, is excluded because it was not possible
to measure a reliable velocity dispersion, which we believe is likely
because there is another nearby clusters in the redshift direction
(\cite{global}).  The reduced $\chi^2$ values of 14 of the 15 clusters
are 1.1 or less, indicating that the NFW function with a fixed scale
radius is a statistically acceptable fit to the density profile.  The
A2390 cluster has $\chi^2=1.38$ per degree of freedom which indicates
that the fit is about 15\% probable.  A2390 is the cluster with the
most data.  If we had about a third less data for this cluster, the
fit would have been entirely acceptable. However, as the amount of
cluster data increases beyond about 100 galaxies, the substructure
within the cluster (\cite{a2390}) becomes more clearly detected. By
extension, if we had obtained more than 200 or so galaxies within any
cluster, the substructure would have become clearly resolved. This
sampling strategy was part of the initial layout of the CNOC
observations.

\section{Discussion and Conclusions}

We have established that the dynamical mass profile, $\rho(r)$, is
proportional to the galaxy number density profile, $\nu(r)$, over the
range of radii where the argument is secure. We further make the
modest assumption that $\nu(r)$ continues to trace the mass, within
the errors, at larger radii. Consequently we conclude that the NFW
function accurately describes the mass field. The measured scale
radius is $0.20\le a_\rho \le 0.30$ with a 95\% confidence range of
$0.12-0.50$. This relatively large confidence range is a result of the
relatively small change in profile slope at the characteristic scale
for the NFW profile. The Hernquist function has much smaller errors
when used in same fitting procedure.  The predicted values for
$\Omega=0.2$ and a CDM spectral normalization of $\sigma_8=0.95$ are
0.20 for an open model (nearly identical to the $\sigma_8=0.6$,
$\Omega=1$ value of 0.19) and 0.26 for a flat low density model
(\cite{nfw2}), are remarkably precise predictions of the observed
value.

Overall it appears that a very good understanding of the evolution of
the mass field in clusters is emerging. The value of $\Omega$
determined from dynamical observations of clusters
(\cite{global,profile}) predicts a slow change in the abundance of
clusters with redshift, which is in accord with the directly observed
value (\cite{s8}).  Previously we found that clusters as individuals
have masses inside \r200\ which are accurately predicted from their
velocity dispersions (\cite{global}). Here we have shown that the
predicted scale radius and the universal NFW profile provide a
statistically acceptable description of the virialized region, $r\le
1.5\r200$, of 14 of the 16 clusters selected for the CNOC
observational program, with the failures likely being the result of
large substructures. The combination of successful predictions of
galaxy cluster mass profiles and their cosmological number density
evolution is a strong argument for the model of cluster formation in
an effectively cold, collisionless, dark matter dominated universe.
These results could be substantially tightened, in the first instance,
through redshift determinations for more cluster galaxies in the
outskirts of this sample.

\acknowledgments
The referee, Cedric Lacey, provided valuable comments which
substantially improved the presentation of this paper.  We thank CFHT
for the technical support which made these observations feasible.
NSERC and the NRC provided financial support.

\clearpage
\begin{table}[h]
\caption{Goodness of Fit of A Universal Profile\label{tab:chi}}
\begin{tabular}{rrr}
Name & $\chi^2/\nu$ & \# cluster redshifts \\
      A2390&  1.38& 178\\
MS0016$+$16&  0.49&  47\\
MS0302$+$16&  0.88&  26\\
MS0440$+$02&  0.62&  37\\
MS0451$+$02&  0.76& 114\\
MS0451$-$03&  0.83&  50\\
MS0839$+$29&  0.98&  45\\
MS1006$+$12&  0.50&  28\\
MS1008$-$12&  0.70&  67\\
MS1224$+$20&  1.10&  24\\
MS1231$+$15&  0.69&  76\\
MS1358$+$62&  0.57& 171\\
MS1455$+$22&  0.69&  55\\
MS1512$+$36&  0.48&  38\\
MS1621$+$26&  0.88&  96\\
\end{tabular}
\end{table}

\clearpage

\figcaption[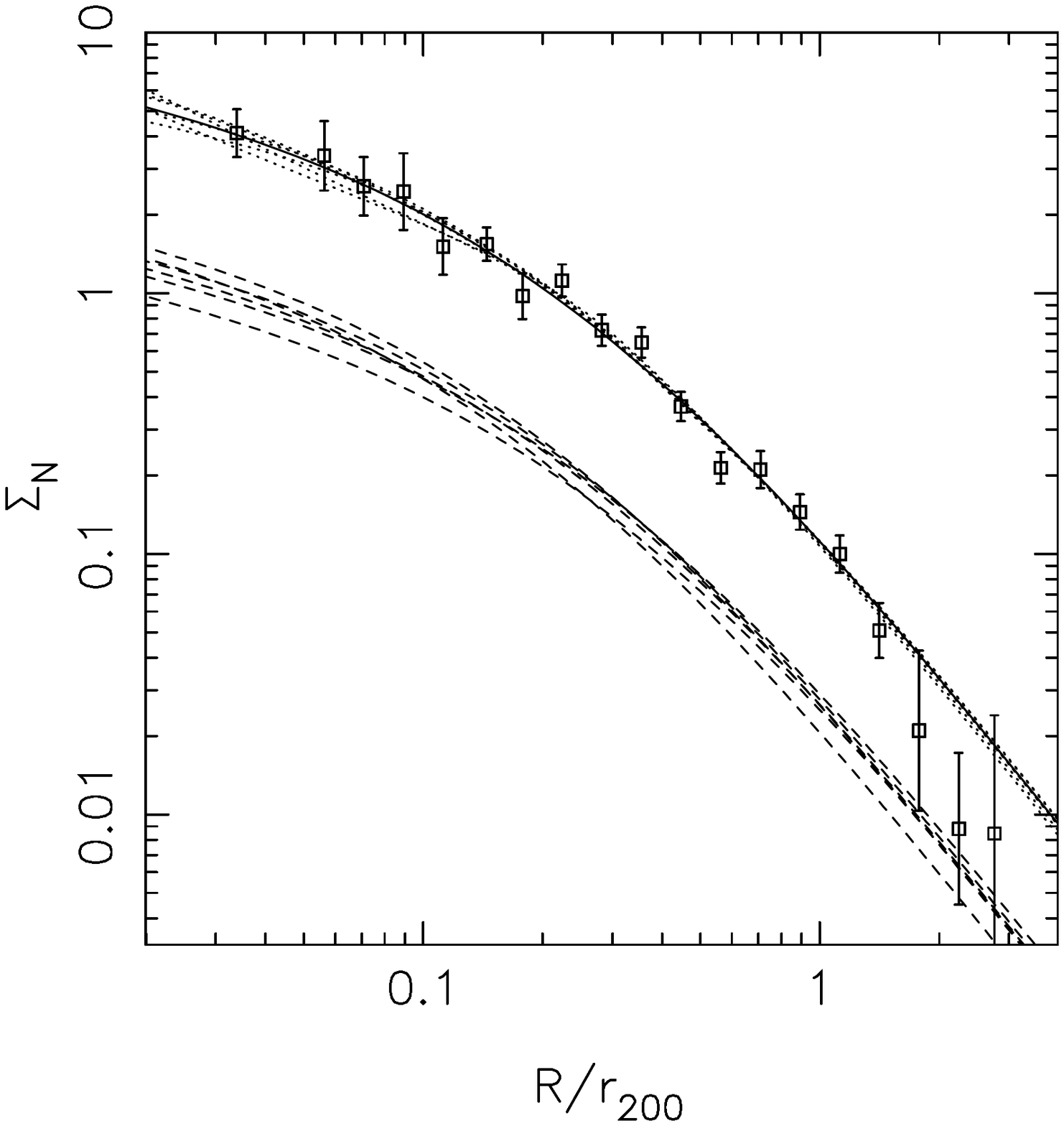]{The average galaxy density profile of
the clusters using the BCG as the center. In the text, we demonstrate
that this galaxy number density profile is statistically equal to the
projected average mass profile. The solid line is the projection of
the NFW function. The dotted lines are the projection of the product
of an assumed NFW {\em mass} profile with the dynamically derived
local light-to-mass ratio, $\nu(r)/\rho(r)$. The dashed lines are the
projected NFW mass profiles themselves, offset downward by half a dex.
\label{fig:surf}}

\figcaption[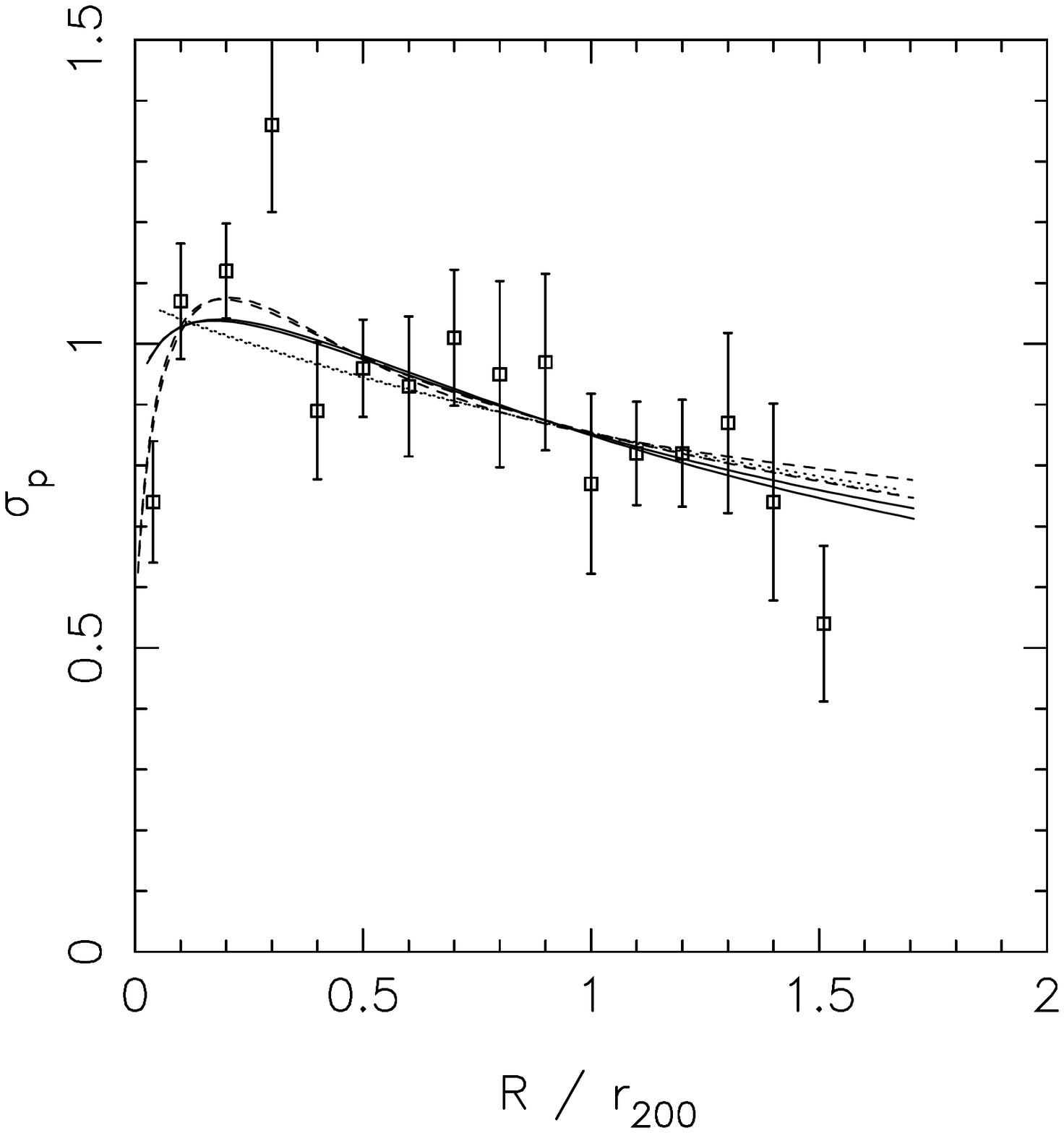]{The projected velocity dispersion profile and
the projection of the fitted profiles, for both $\beta_m=0.3$ and
$0.5$ which are nearly indistinguishable on this plot. The dotted line
is for $[c_1,c_2]=[0,1]$, the dashed for $[1,0]$ and the solid is
$[8,1/2]$ (see text for details).
\label{fig:sig}}

\figcaption[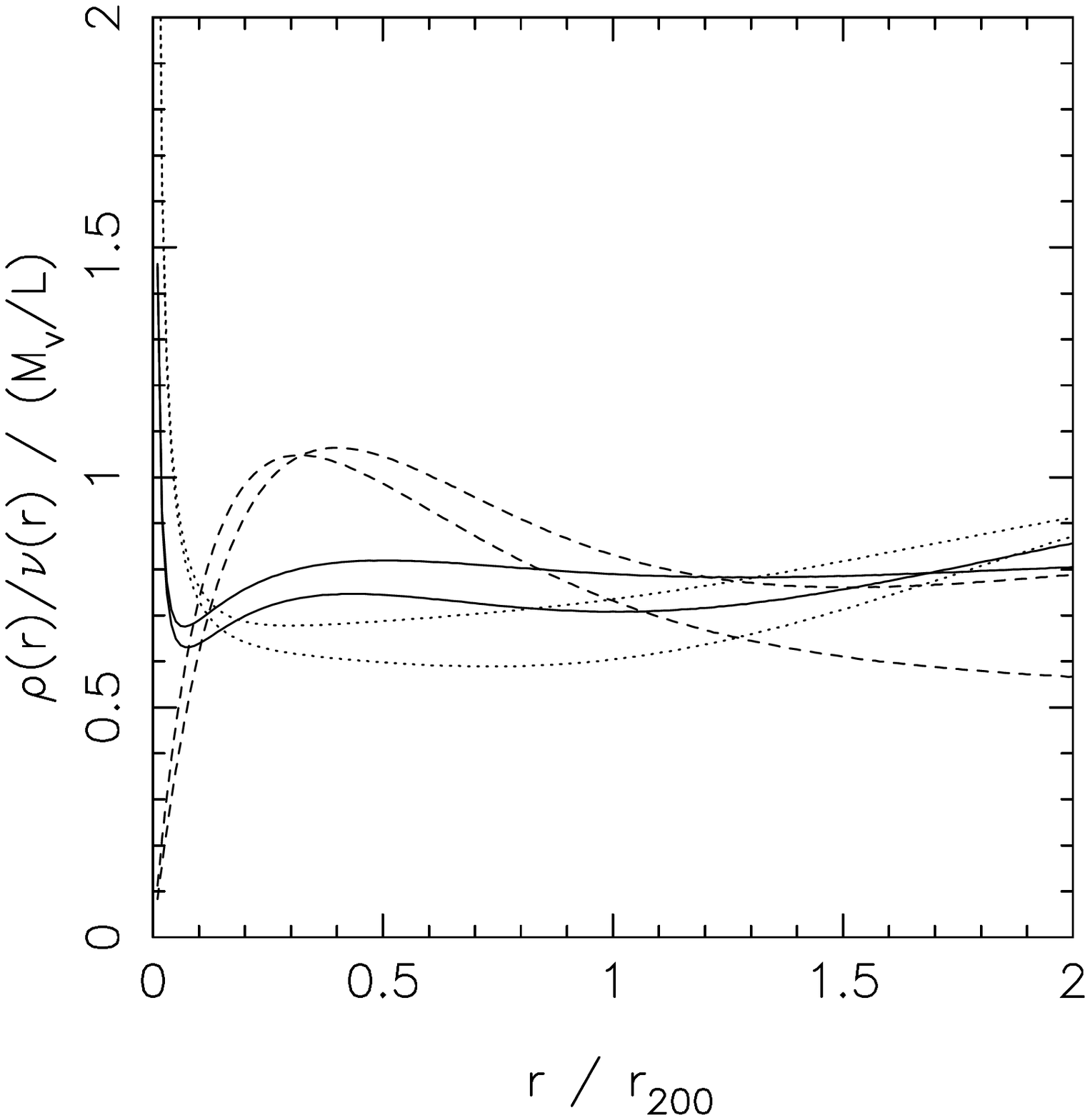]{The derived ratio of the dynamical mass density profile,
$\rho(r)$, to $r$ selected galaxy profile, $\nu(r)$, normalized with
the virial mass-to-light ratio evaluated inside $1.5\r200$.  In each
pair of curves the upper line at small radius is for $\beta_m=0.3$ and
the lower for $\beta_m=0.5$. The dotted line is for $c_1=0$, the
dashed for $c_2=0$ and the solid is our preferred $c_1=8$, $c_2=1/2$.
\label{fig:dol}}

\begin{figure}[h] \figurenum{1}\plotone{fig1.ps} \caption{}\end{figure}  
\begin{figure}[h] \figurenum{2}\plotone{fig2.ps} \caption{}\end{figure}  
\begin{figure}[h] \figurenum{3}\plotone{fig3.ps} \caption{}\end{figure}  

\end{document}